\documentclass[sigconf, screen, nonacm]{acmart}

\usepackage{multicol}
\usepackage{multirow}
\usepackage{array}
\usepackage{tikz}
\usepackage{pgfplots}

\AtBeginDocument{%
  }

\begin{document}

\title{BigPower: Hierarchical Source-Level Module Power Estimation for CPUs with Large Language Models}


\author{Honghua Zhu}
\affiliation{%
  \institution{State Key Lab of Processors, Institute of Computing Technology, Chinese Academy of Sciences}
  \city{Beijing}
  \country{China}
}
\affiliation{%
  \institution{University of Chinese Academy of Sciences}
  \city{Beijing}
  \country{China}
}
\email{zhuhonghua22s@ict.ac.cn}

\author{Chunjie Luo} 
\authornote{The corresponding author is Chunjie Luo (\textit{luochunjie@ict.ac.cn}).}
\affiliation{%
  \institution{State Key Lab of Processors, Institute of Computing Technology, Chinese Academy of Sciences}
  \city{Beijing}
  \country{China}
}
\email{luochunjie@ict.ac.cn}

\author{Jianfeng Zhan}
\affiliation{%
  \institution{State Key Lab of Processors, Institute of Computing Technology, Chinese Academy of Sciences}
  \city{Beijing}
  \country{China}
}
\email{zhanjianfeng@ict.ac.cn}

\fancyhead{}
\renewcommand\shortauthors{}



\begin{abstract}

Accurate power estimation is important for understanding and optimizing CPU power behavior, yet practical workflows often rely on simulation-derived information or post-silicon analysis. In this work, we present BigPower, a hierarchical source-level surrogate model for fine-grained module-level power estimation during CPU design. BigPower leverages large language model-based representations together with architectural hierarchy, module connectivity, configuration parameters, and workload context to estimate module-level power consumption directly from source-level design information, without requiring additional simulation during inference. Experimental results in the open-source XiangShan processor family demonstrate practical fine-grained power estimation across diverse configurations and workloads, offering an efficient alternative to conventional simulation-based workflows.

\end{abstract}

\maketitle

\section{Introduction}

Improving CPU power efficiency has become increasingly important as modern computing workloads continue to grow in scale and complexity.

There are three objectives for CPU power consumption evaluation: 1) Accuracy. Precise evaluation is critical, as inaccuracies can significantly misguide CPU design decisions. 2) Cost. Minimizing evaluation cost, particularly time, is essential to accelerate the iteration cycle of CPU design optimization. 3) Granularity. This encompasses both spatial and temporal dimensions. Spatial granularity refers to whether the estimation is performed on the entire chip or on specific modules within it. In terms of temporal granularity, per-cycle power estimation is a research hot spot.

Gate-level power analysis is generally considered the most accurate approach for power sign-off, which utilizes power-characterized standard cell libraries with commercial tools such as Synopsys PrimeTime PX (PTPX) by feeding in gate-level netlists and simulation results. RTL-level power modeling tools, such as PowerArtist, PowerPro, Apollo \cite{Xie2021APOLLOAA}, Primal \cite{2019ZhouPRIMALPowerInference}, and PACE \cite{2024BenarPACEMLPBasedFast}, utilize RTL simulation traces as training features for power consumption estimation, but their accuracy is relatively lower. Positioned between gate-level and RTL-level modeling, GRANNITE \cite{2020ZhangGRANNITEGraphNeural} utilizes two  inputs: a graph representation of the gate-level netlist and dynamic toggle rates of input ports and register outputs obtained via RTL simulation.

Architecture-level models, such as CACTI \cite{li2011cacti}, Wattch \cite{brooks2000wattch}, McPAT \cite{2009LiMcPATIntegratedPower}, McPAT-PVT \cite{Tang2015McPATPVTDA}, McPAT-Monolithic \cite{Guler2020McPATMonolithicAA}, McPAT-Calib \cite{2023ZhaiMcPATCalibRISCVBOOM}, and FirePower \cite{2025ZhangFirePowerFoundationGeneralizable}, typically trade estimation accuracy for lower evaluation cost and higher scalability. In terms of evaluation cost, however, the opposite is true: the time cost increases sequentially for architecture-level, RTL-level, and gate-level evaluations. Additionally, RTL-level and gate-level evaluations allow for fine-grained power modeling, while architecture-level modeling tends to be relatively coarse-grained.

Consequently, many practical power estimation workflows still rely on simulation-derived activity information, such as architectural event statistics or RTL/gate-level switching activities, to construct and calibrate power models. The finer the simulation granularity, the more reliable and detailed the captured information becomes, leading to higher accuracy and finer-grained modeling. However, this comes at the expense of increased time costs, which may slow iterative CPU optimization and evaluation.

Besides, CPU design space exploration requires power models to be capable of making predictions for unseen designs. In contrast, frameworks such as Primal \cite{2019ZhouPRIMALPowerInference} and Apollo \cite{Xie2021APOLLOAA} are only modeled for fixed designs; when the design changes, the models must be reconstructed and retrained. While McPAT-Calib \cite{2023ZhaiMcPATCalibRISCVBOOM} utilizes CPU configuration parameters and dynamic features generated by the Gem5 simulator to provide predictions for new CPU designs, it lacks the capability for fine-grained predictive analysis of the thousands of individual modules within the CPU. GRANNITE \cite{2020ZhangGRANNITEGraphNeural} leverages Graph Neural Networks (GNNs) to perform transferable predictions of gate-level switching activity in combinational circuits. However, GRANNITE requires RTL simulation to obtain dynamic waveforms and necessitates logic synthesis to construct the connectivity graph of all constituent gates.

In this paper, leveraging large language models (LLMs), we propose BigPower, a model that enables fine-grained, module-level power consumption prediction directly at the source code level without the need for simulation. This allows us to identify power consumption bottlenecks in fine-grained modules at a lower time cost and with greater flexibility, thereby accelerating the cycle of design optimization.

Our motivation is based on the observation that CPU power consumption is shaped by workload characteristics, microarchitectural configurations, and hierarchical module interactions. Existing approaches typically obtain such information through simulation-derived activity traces, which incur substantial cost. However, many relevant structural signals—including RTL implementation, module hierarchy, connectivity, architectural configurations, and workload descriptions—are already available at the source level. We hypothesize that learning from these heterogeneous signals can enable practical module-level power estimation without requiring additional simulation during inference. Large language models provide a flexible framework for encoding long-form structured source information and heterogeneous contextual representations, making them a promising foundation for this task.

To build BigPower, we first label a large module-level power dataset for training purposes. Utilizing seven different workloads, we conduct power labeling through EDA-flow (Electronic Design Automation) across 400 microarchitecture configurations of the open-source XiangShan processor family \cite{xu2022towards}. With approximately 5,000 module instances in XiangShan, the resulting module-level power labeling data amount to nearly $  1.4\times 10^{7} $ entries.

Next, we fine-tune our BigPower model based on an open-source pre-trained large language model. Constructing the input for the large language model is crucial. The input, denoted as $x=(c,w)$, comprise workload information ($w$) and a collection of contextual data ($c$) that represents CPU information. The corresponding output $y$ represent the power consumption values of the module instances. The most significant challenge lays in representing the CPU information. We aim for the large model to comprehend both the overall CPU and the specific module instances requiring power prediction. However, due to GPU memory limiting, the input text length for large models and the complexity of the CPU code, it is impractical to input the entire CPU code into the model. Therefore, we propose a CPU information representation structure that combines global and local information, including microarchitecture configurations, CPU hierarchical structure information, module connectivity information, and module RTL code. The microarchitecture configurations comprise 20 microarchitecture parameter values which would be varied. The hierarchical structure of the CPU code could be viewed as a tree, and the hierarchical information we utilize included the path from the top-level module to the current instance, the sub-modules contained within the current module, and their power consumption values. The module connectivity information encompasses the names of all modules connected to the current module via signal lines. Given the varying lengths of different module RTL codes, we divide long code into several code segments, each constructing a separate input $x$. After constructing the $(x,y)$ pairs, we employ these pairs to fine-tune the pre-trained open-source large model in a supervised manner.

Finally, we utilize ensemble learning and a bottom-up hierarchical approach to predict module-level power consumption. During prediction, we encounter two issues: 1) During fine-tuning, our hierarchical structure information include the power consumption of sub-modules. Therefore, we adopt a bottom-up prediction approach, first predicting the power consumption of bottom-level modules and then progressively predicting the power consumption of upper-level modules layer by layer. 2) Modules with long Verilog code have multiple code segments, each constructing an $x$ and yielding multiple prediction values. We average these multiple prediction values, which is an idea of ensemble learning. That also mitigates the randomness of large language model predictions.

We test the within-family generalization ability of BigPower in predicting power consumption under new CPU configurations, new workloads, and combinations of both. BigPower achieves $R^2>0.9$ prediction performance across these tests, demonstrating its capability to perform fine-grained, module-level power consumption predictions at the source code level.

Our contributions are as follows:
\begin{itemize}

\item We propose and implement fine-grained, module-level CPU power consumption prediction at the source code level without requiring simulation during inference, enabling us to identify power consumption bottlenecks in fine-grained modules at a lower time cost and with greater flexibility.

\item We develop an LLM-based framework for CPU power consumption prediction, constructed a 14 million-entry power consumption dataset for training the LLM-based power model, and validate its effectiveness on the open-source XiangShan processor family.

\item We introduce a CPU information representation structure that combines global and local information to address the challenge of directly inputting comprehensive CPU code information into large language models.

\item We propose an ensemble learning and bottom-up hierarchical prediction method for module-level power consumption prediction.

\end{itemize}

\begin{figure}[h!]
\centering
\scalebox{0.55}{
\begin{tikzpicture}
    \begin{axis}[
        clip=false,
        width=14cm,
        height=10cm,
        x axis line style = {->},
        xlabel style={at={(ticklabel* cs:1.02)}, anchor=west}, 
        xmin=0.5,
        xmax=4.3,
        xtick={ 0,1,2,3,4,5}, 
        xticklabels={ , Library-based, Regression, ML-based,LLM-based}, 
        x tick label style={text width=2.5cm, align=center},
        y axis line style = {->}, 
        ylabel style={at={(ticklabel* cs:1.05),}, anchor=south}, 
        ymin=0.5,
        ymax=4.5,
        ytick={0, 1, 2, 3, 4,5}, 
    yticklabels={, Gate Level, RTL Level, Architecture Level,Source Code Level}, 
        enlarge x limits=0.15,
        enlarge y limits=0.15,
        tick style={draw=none}, 
        axis lines=left,
        ]
        \addplot[only marks, mark=*, mark size=3pt] coordinates {(1,1)};
        \node[below right=1pt] at (axis cs:1,1) {PrimeTime PX};
        \addplot[only marks, mark=*, mark size=3pt] coordinates {(2.93,1)};
        \node[above] at (axis cs:2.93,1) {CAPEDL \cite{2024LiuCAPEDLCycleAccuratePower}};

        \addplot[only marks, mark=*, mark size=3pt] coordinates {(0.9,2)};
        \node[above] at (axis cs:0.9,2) {PowerArtist};
        \addplot[only marks, mark=*, mark size=3pt] coordinates {(1.1,2)};
        \node[below] at (axis cs:1.1,2) {Joules};
        \addplot[only marks, mark=*, mark size=3pt] coordinates {(2.5,3)};
        \node[below left] at (axis cs:2.5,3) {PANDA \cite{2023ZhangPANDAArchitectureLevelPower}};
        \addplot[only marks, mark=*, mark size=3pt] coordinates {(2.6,3)};
        \node[above left] at (axis cs:2.6,3) {FirePower \cite{2025ZhangFirePowerFoundationGeneralizable}};
        \addplot[only marks, mark=*, mark size=3pt] coordinates {(1.9,2)};
        \node[above left] at (axis cs:1.9,2) {Simmani \cite{Kim2019SimmaniRP}};
        \addplot[only marks, mark=*, mark size=3pt] coordinates {(2,2)};
        \node[below=2pt] at (axis cs:2,2) {Apollo \cite{Xie2021APOLLOAA}};
        \addplot[only marks, mark=*, mark size=3pt] coordinates {(2.1,2)};
        \node[above ] at (axis cs:2.1,2) {DEEP \cite{Xie2022DEEPDE}};
        \node[left] at (axis cs:2.95,1.5) {GRANNITE \cite{2020ZhangGRANNITEGraphNeural}};
        \addplot[only marks, mark=*, mark size=3pt] coordinates {(3,1.5)};
        \addplot[only marks, mark=*, mark size=3pt] coordinates {(3,2)};
        \node[above left] at (axis cs:3.1,2) {PRIMAL \cite{2019ZhouPRIMALPowerInference}};
        \addplot[only marks, mark=*, mark size=3pt] coordinates {(3.1,2)};
        \node[below right] at (axis cs:3.1,2) {PACE \cite{2024BenarPACEMLPBasedFast}};
        \addplot[only marks, mark=*, mark size=3pt] coordinates {(3.2,2)};
        \node[above right] at (axis cs:3.2,2) {MasterRTL \cite{Fang2025TransferablePP}};
        \addplot[only marks, mark=*, mark size=3pt] coordinates {(1,3)};
        \node[above=2pt] at (axis cs:1,3) {McPAT \cite{2009LiMcPATIntegratedPower,2013LiMcPATFrameworkMulticore}};
        \addplot[only marks, mark=*, mark size=3pt] coordinates {(2.9,3)};
        \node[below] at (axis cs:2.85,3) {PowerGear\cite{2022LinPowerGearEarlyStagePowera}};
        \addplot[only marks, mark=*, mark size=3pt] coordinates {(3,3)};
        \node[above=2pt] at (axis cs:3,3) {McPAT-Calib \cite{2023ZhaiMcPATCalibRISCVBOOM}};
        \addplot[only marks, mark=diamond*, mark size=3pt] coordinates {(4,4)};
        \node[above] at (axis cs:4,4) {Our work};
    \end{axis}
\end{tikzpicture}}
\caption{Power consumption analysis landscape. For clarity and simplicity, some related works have been omitted. Notably, Source Code Level implies that predictions are derived directly from static code properties, as opposed to features obtained from dynamic simulation. }
\label{fig:power_tools}
\end{figure}
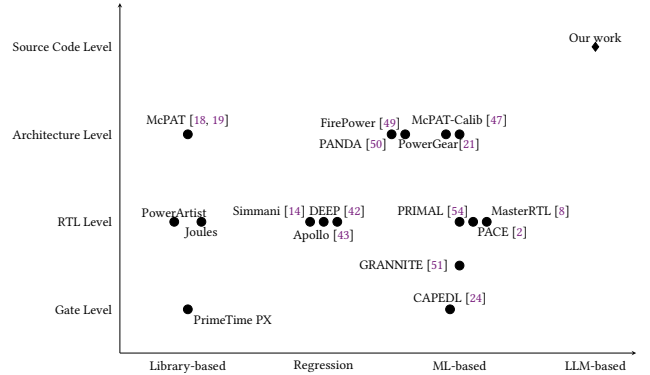

\section{Related Works}

\subsection{Power Analysis and Prediction}
\textbf{Analytical and library-based models:} In order to get high accuracy and corporate with other EDA software, commercial tools for power analysis are usually based on power-characterized standard cell libraries and use analytical methods to produce stable and traceable results. Synopsys PrimeTime PX (PTPX), which is one of the most commonly-used tools and often considered as some sort of golden standard, needs netlist and gate-level simulation waveforms to give precise and fine-grain power estimates. To avoid the time-consuming EDA flow to generate netlist and speed-up power feedback, other tools such as Ansys PowerArtist and Cadence Joules provide a course-grain power estimate using RTL codes and simulations. This rapid feedback capability is crucial for supporting agile hardware development methodologies.

As for academia, McPAT \cite{2009LiMcPATIntegratedPower,2013LiMcPATFrameworkMulticore} is the first to propose an integrated power, area, and timing modeling framework and has been constantly upgraded to adapt to new integrated circuit fabrication technology. McPAT-PVT \cite{Tang2015McPATPVTDA} introduces a new FinFET design library to it and considers influence of process, supply voltage, and temperature (PVT) variations. McPAT-monolithic \cite{Guler2020McPATMonolithicAA} enables modeling hybrid monolithic multi-core architectures. McPAT-7nm, as a part of McPAT-Calib \cite{2023ZhaiMcPATCalibRISCVBOOM} becomes an upgraded version adapted for 7nm technology.

\textbf{Regression models:} Compared to analytical methods, regression-based power models has been extensively researched since linear regression models and its numerous variations are easy to implement and have solid theoretic background. Researchers have developed numerous feature selection algorithms to improve their accuracy and to reduce overhead. Simmani \cite{Kim2019SimmaniRP} constructs a toggle-pattern for signals clustering and then selects these signals to serve as features. Using a minimax concave penalty (MCP)-based feature selection algorithm, Apollo \cite{Xie2021APOLLOAA} utilizes $\le 0.05\%$ of RTL signals to train the
power model using linear regression. DEEP \cite{Xie2022DEEPDE} proposes a bit-level RTL signal selection and a two-step on-chip power meters selection method. The work \cite{2024StuckmannEnergyAwareRegisterAllocation} uses an interaction linear regression power model in order to guide register allocation. The literature \cite{2025TavakkoliSensorlessPowerMeasurement} proposes that using CMOS MEMS magnetic field sensors combined with some commonly monitored system parameters, a multivariate nonlinear regression is capable to predict system power consumption well.

\textbf{Machine learning model:} Recent development of machine learning models
enables capturing complex and intricate relationships in power consumption. Kumar et al. \cite{2023KumarMachineLearningBasedMicroarchitecture} build up component level power model, testing and comparing multiple ML regressors such as decision tree, gradient boosting model and random forest. In McPAT-Calib \cite{2023ZhaiMcPATCalibRISCVBOOM}, XGBoost regressor serves to calibrate the output of McPAT. PANDA \cite{2023ZhangPANDAArchitectureLevelPower} develops its own simpler analytical function for each component and then uses XGBoost to handle more complex patterns. FirePower \cite{2025ZhangFirePowerFoundationGeneralizable} introduces separate hardware and event models, both based on XGBoost, to adapt to few-shot learning scenario.

In recent years, neural networks have also been applied to this domain. PRIMAL \cite{2019ZhouPRIMALPowerInference} uses a convolutional neural network(CNN) to model cycle-by-cycle RTL power with detailed switching activities of all registers. GRANNITE  \cite{2020ZhangGRANNITEGraphNeural} utilizes graph neural networks(GNNs) to predict gate-level toggle rates. Also, with GNN, PowerGear \cite{2022LinPowerGearEarlyStagePowera} predicts the power of FPGA in the high-level synthesis stage. MasterRTL \cite{Fang2025TransferablePP} develops a new bit-level design representation and the power model part of it uses either a GNN or a lightweight tree-based model. It is worth noticing that an LLM-based data augmentation is introduced. The work \cite{Zhai2023MicroarchitecturePM} uses a multilayer perceptron (MLP) with residual connection as a power model and develops a cross-domain mix-up data generation for transfer learning. In CAPEDL \cite{2024LiuCAPEDLCycleAccuratePower}, a two-step deep learning network is introduced, including an auto-encoder for signal compression and a MLP for ASIC designs' cycle-accurate power estimation. PACE \cite{2024BenarPACEMLPBasedFast} uses a sparse Multilayer Perceptron (sMLP) on the selected register bits to predict the power consumption of the CPU designs. 

All works mentioned above need simulations to extract input features, whether architecture level  \cite{2023ZhaiMcPATCalibRISCVBOOM,2023ZhangPANDAArchitectureLevelPower,2023KumarMachineLearningBasedMicroarchitecture,2022LinPowerGearEarlyStagePowera,2025ZhangFirePowerFoundationGeneralizable,Zhai2023MicroarchitecturePM}, RTL level \cite{2019ZhouPRIMALPowerInference,2020ZhangGRANNITEGraphNeural,2024BenarPACEMLPBasedFast,Fang2025TransferablePP} or gate level \cite{2024LiuCAPEDLCycleAccuratePower}. The requirement of simulations draws some disadvantage. Designs in the early stage are frequently incomplete or contain errors, rendering direct simulation infeasible without further refinement. Further more, relying on simulations introduces unnecessary correlation between different teams, because some specific and localized bugs might have negligible influence on other modules' power optimization.

\subsection{Large Language Models}

The Transformer architecture, introduced in 2017  \cite{vaswani2017attention}, fundamentally shifted neural network design for natural language processing. By entirely replacing recurrence with a self-attention mechanism, it enabled superior parallel processing and long-range dependency handling, drastically improving training efficiency and model capacity. This led to the pervasive pre-training and fine-tuning paradigm. Subsequent models like GPT \cite{radford2018improving} and its successors (GPT-2 \cite{Radford2019LanguageMA}, GPT-3 \cite{Brown2020LanguageMA}) advanced large-scale autoregressive pre-training, showcasing remarkable generation and pioneering in-context learning. Further, InstructGPT \cite{ouyang2022training} focused on aligning LLMs with human preferences. More recently, the release of open-source models such as LLaMA \cite{touvron2023llama} and Qwen \cite{Bai2023QwenTR} has broadened access to LLM technology, fostering wider research and development. 

Large Language Models are increasingly being investigated for their capabilities within the domain of computer architecture. Current research demonstrates their utility in various aspects, including RTL code generation \cite{Chang2024DataIA, zhao2024codev, liu2024rtlcoder, chen2025chipseek}, test bench creation and verification \cite{Huang2024TowardsLV, ma2024verilogreader}, and even entire chip design \cite{wang2024chatcpu,firouzi2025chipmnd}. To support advancements in LLM-driven code generation, specialized benchmarks such as VerilogEval \cite{liu2023verilogeval} and RTLLM \cite{lu2024rtllm} have been established. Furthermore, LLMs are being actively adopted for High-Level Synthesis (HLS) code generation \cite{Xiong2024HLSPilotLH, gai2025exploring} and optimization tasks  \cite{mashnoor2025timelyhls}. LLMs also present capabilities in design space exploration \cite{wang2025llm,xu2025intelligent4dse,tang2025chatdse}.

In this paper, we utilize large language models to predict power consumption, enabling fine-grained, module-level power consumption prediction directly at the source code level without the need for simulation. Fig. \ref{fig:power_tools} demonstrates the differences between our work and other works. Our proposed approach offers greater flexibility in power estimation, enabling rapid power analysis of arbitrary modules within new designs.

\section{Methods}

We frame the power estimation task as a supervised learning problem. The goal is to learn a function, $f$, that maps an input pair, $x=(c,w)$, to its corresponding power consumption, $y$. Here, $c$ represents the comprehensive instance information, encompassing details such as its Verilog code, the CPU context it belongs to, and other relevant design parameters. $w$ represents the software workload. The model's objective is to approximate this function:
\begin{equation}
    y=f(x)=f(c,w)
\end{equation}

\begin{figure}[!htbp]
\centering
\includegraphics[width=1\columnwidth]
{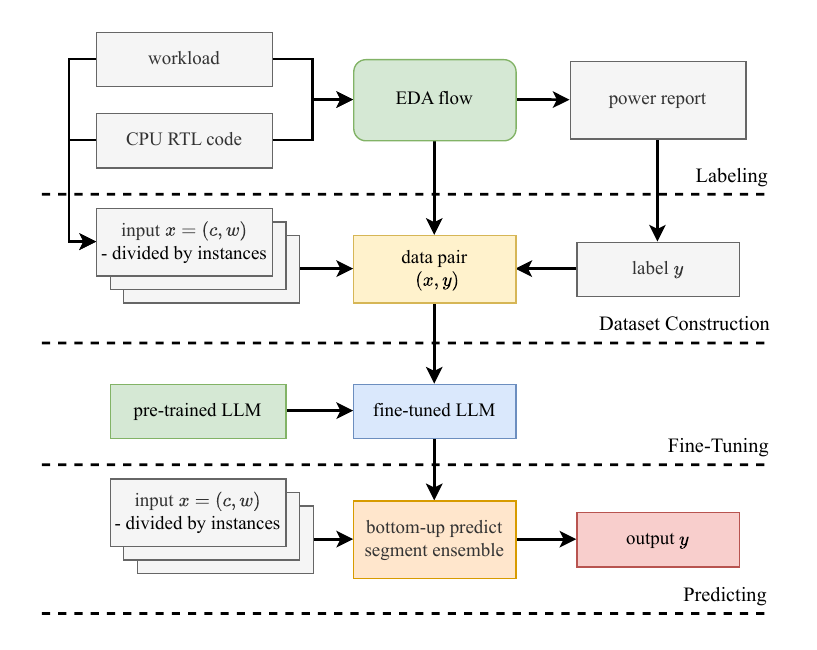}
\caption{The overview of BigPower.}
\label{fig-power-model}
\end{figure}

Currently, there are numerous open-source pre-trained large language models available within the community. Adapting these general-purpose models to a specific vertical domain lies in the construction of a specialized dataset, which is then used to fine-tune the base model for domain-specific tasks. Therefore, for constructing BigPower, as illustrated in the Fig. \ref{fig-power-model}, the critical steps are as follows:

\begin{itemize}
\item Labeling. Acquiring and assigning ground-truth values of power consumption $y$ to certain instances of $x$.
\item Dataset construction. The primary focus here is on how to represent $x=(c,w)$ and then pair it with $y$ to form training pairs $(x,y)$.
\item Fine-tuning. Utilize the $(x,y)$ pairs to fine-tune the pre-trained large language model to adapt it to the power analysis domain.
\item Predicting. Perform power consumption predictions for new workloads or new CPU modules.
\end{itemize}

Next, we will begin by introducing the acquisition of ground truth power consumption values $y$ in Section \ref{sec-labeling}. Subsequently, in Section \ref{sec-input}, we will present the complete representation of $x=(c,w)$. Section \ref{sec-fine} will introduce the fine-tuning process, and finally, Section \ref{sec-predicting} will explain how to conduct predictions.

\subsection{Labeling} \label{sec-labeling}

We label the power consumption of different instances of open-source XiangShan processor family \cite{xu2022towards}, by running various workloads on XiangShan with different microarchitecture configurations. 

We use seven workloads which come from commonly used CPU benchmark suites. They include 
\begin{enumerate}
  \item hanoi: hanoi tower.
  \item kmeans: k-means clustering algorithm.
  \item qsort: quick sort algorithm.
  \item multiply: multiplication implemented by  bit-wise operation.
  \item fib: matrix-multiplication-based fast power to calculate Fibonacci sequence.
  \item mm\_int: matrix multiplication of integer data type.
  \item mm\_float: matrix multiplication of single-floating-point data type.
\end{enumerate} 

The microarchitecture configurations are sampled from the space defined by Table \ref{designspace}, which encompasses approximately $5.8 \times 10^9$ possible configurations. We perform power labeling across nearly 400 microarchitecture configurations of the XiangShan. With approximately 5,000 module instances in XiangShan, the resulting module-level power labeling data amount to nearly $  7\times 400 \times 5000 = 1.4\times 10^{7} $ entries.

\begin{table}[!htbp]
    \caption{Microarchitecture configuration space. }
    \label{designspace}
    \centering
    \begin{tabular}{c|c}
        \hline
          Name &  Candidate Value\\
        \hline
          ras\_entry & 8,16,24,32 \\
          decode\_width &  4,6  \\
          rob\_entry &  64,128,192,256,320\\
          phy\_register &  64,128,192,256,320 \\
          int\_multdiv &  1,2,3,4\\
          fp\_misc &  2,4,6,8  \\
          fp\_mac &  2,4,6,8\\
          ldq\_entry &  32,48,64,80,96 \\
          stq\_entry &  32,48,64,80,96\\
          icache\_way &  2,4,8\\
          icache\_tlb &  16,32,48,64,80 \\
          icache\_size (KB) &  32,64,128 \\
          dcache\_way &  2,4,8  \\
          dcache\_tlb &  32,48,64,80,96 \\
          dcache\_size (KB) &  32,64,128\\
          dcache\_mshr &  16,32,64\\
          l2\_size (MB) &  1,2,4\\
          l3\_size (MB) &  2,4,8,16,32 \\
          l2\_assoc &  4,8,16\\
          l3\_assoc &  4,8,16 \\

        \hline
    \end{tabular}
\end{table}

\subsection{Dataset Construction} \label{sec-input}

Large language models are based on Transformer \cite{vaswani2017attention} architecture. The attention mechanism of traditional Transformers exhibits a quadratic time complexity ($O(n^2)$) with respect to the input sequence length $n$, posing inherent challenges for long context processing. Although modern LLMs have seen advancements in extending context windows, the practical deployment of very large context lengths remains constrained by our limited memory and computational resource budgets. Therefore, it is necessary to partition the comprehensive power estimation task for an entire CPU design into distinct instances. The question is how. 

\subsubsection{Overview}

\begin{figure*}[!htbp]
\centering
\includegraphics[width=2\columnwidth]
{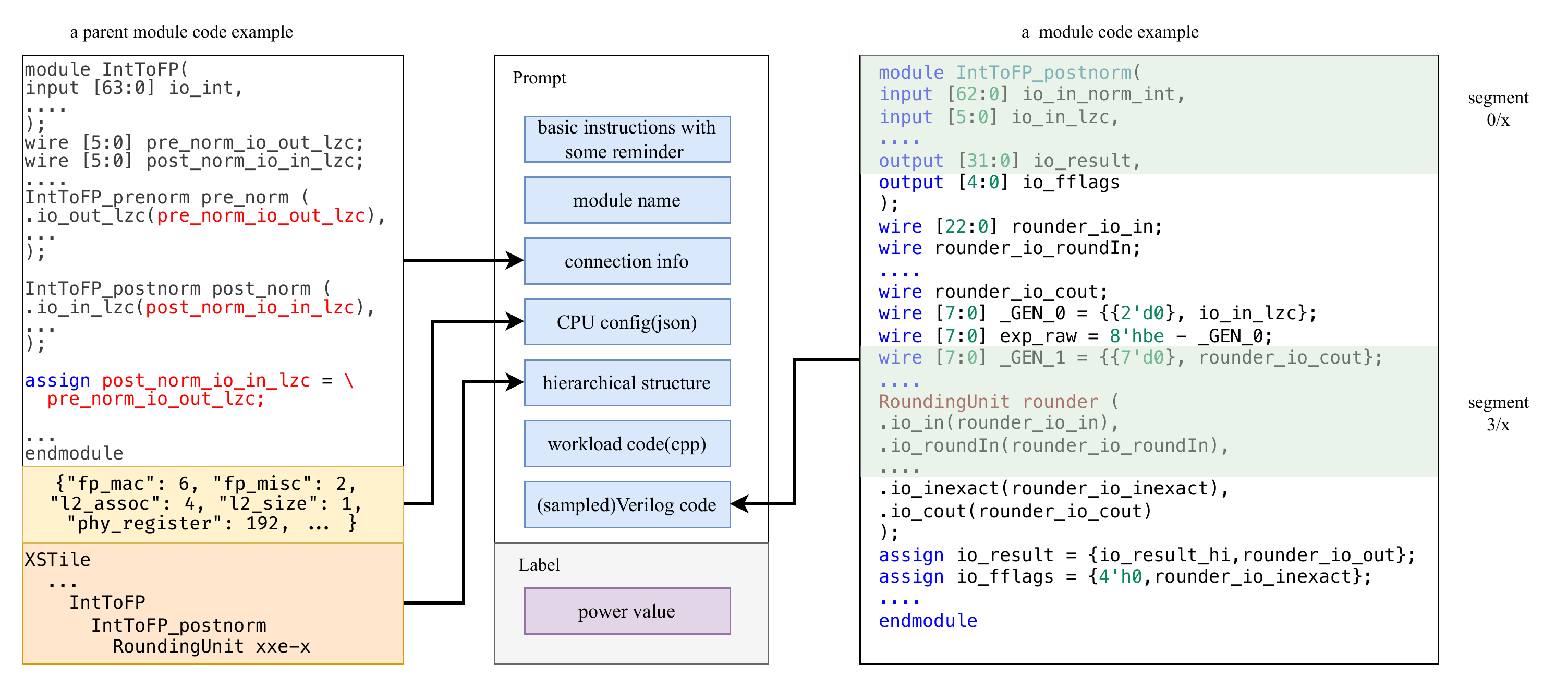}
\caption{Comprehensive illustration of data pair construction for power prediction. The figure provides a detailed view of the input $x$ (comprising a prompt structure and associated data) and the output $y$ (power value). The central panel illustrates the composition of the LLM prompt. The right panel shows a \texttt{IntToFP\_postnorm} module's partial Verilog code, highlighting how it is divided into segments for input. The upper-left panel depicts a parent module \texttt{IntToFP} and its two interconnected child module instances (\texttt{IntToFP\_prenorm, IntToFP\_postnorm}), while the middle-left panel presents CPU configuration parameters in JSON format. The bottom-left panel depicts the hierarchical structure of the \texttt{IntToFP\_postnorm} module. 
}
\label{fig-represent-model}
\end{figure*}

Consider a hypothetical scenario: a fine-tuned LLM achieves power consumption prediction accuracy comparable to commercial EDA tools. This thought experiment prompts us to investigate the theoretical minimum information required for such an achievement. 

Traditional commercial EDA flow typically need standard cell libraries, RTL code or netlist and simulation waveforms. When developing a CPU within a specific technology node, standard cell libraries can be considered static and thus learned as background knowledge by the model. Our work focuses on front-end factors affecting power consumption, which are independent of the later-stage physical implementation process. To isolate our analysis, we assume a constant, high-quality physical implementation, effectively neutralizing the impact of these typical back-end factors.

However, the power consumption of an individual module instance is intrinsically linked not only to the instance's own characteristics but also to the broader context of the entire CPU design and its interactions with other modules. Consequently, the isolated Verilog definition of an instance (as discussed in Section \ref{sec-sample-verilog}) is insufficient. To address this, connection information (detailed in Section \ref{sec-conn-info}) and the hierarchical structure of the design (elaborated in Section \ref{sec-hier}) are crucial inputs. Furthermore, for a given design, simulation waveforms are determined by the applied workload, a topic explored in Section \ref{sec-workload}.

In summary, the complete input pair, $x=(c,w)$, is illustrated in Fig. \ref{fig-represent-model}. These processing steps are minor, typically completing within a few minutes. Our method requires only partial information from the RTL design and thus is much faster than compilation and elaboration step necessary for conventional simulation, which must parse and comprehend the entire design's semantic and behavioral logic to generate executable code.

\subsubsection{Connection Information}\label{sec-conn-info} The upper left side of Fig. \ref{fig-represent-model} shows an example that one parent module \texttt{IntToFP} has two instantiated child modules \texttt{IntToFP\_prenorm} and \texttt{IntToFP\_postnorm}. It is worth noticing that there are two red-colored signals, \\
\texttt{pre\_norm\_io\_out\_lzc} and \texttt{post\_norm\_io\_in\_lzc}, mapped to child modules' input or output ports and they have direct connection to each other by the blue-colored Verilog keyword \texttt{assign}. The \texttt{assign} statement in Verilog facilitates direct continuous assignment compared with procedural assignment. We consider that any two instances have interaction with each other only if they have common signals identified by \texttt{assign} statement. A Python binding of slang (Pyslang)  \cite{Slang}, which is a fast and compliant SystemVerilog frontend, is used to extract those relationship. This information will be cached to speed up dataset construction and later be represented in a natural language tongue. 

\subsubsection{Hierarchical Structure} \label{sec-hier}
\begin{figure}[!htbp]
\centering
\includegraphics[width=1.0\columnwidth]
{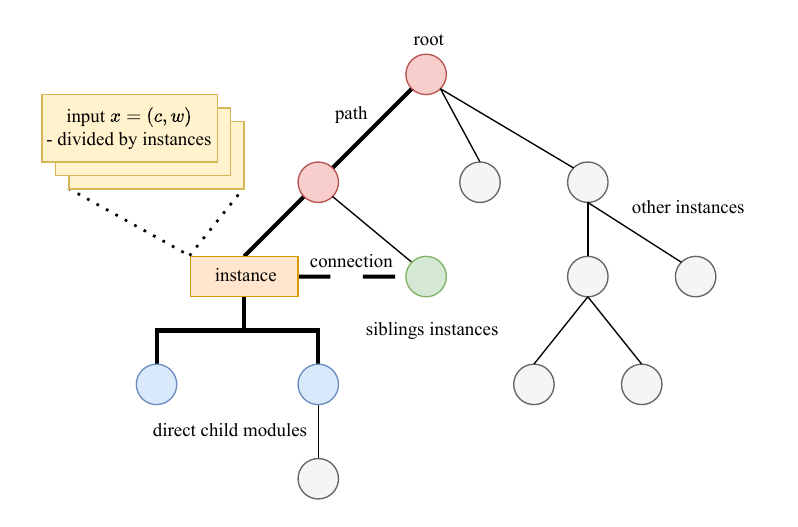}
\caption{Hierarchical tree representation of a CPU design. Each node (circles and rectangle) signifies a module instance. An example instance (orange) is highlighted, showing how the input $x=(c,w)$ is constructed per-instance. The diagram also depicts hierarchical relationships (e.g., 'root', 'path', 'direct child modules') and dotted-line inter-instance 'connection' (between siblings).}
\label{fig-cpu-tree}
\end{figure}

The right side of Fig.  \ref{fig-represent-model} is a demonstration of Verilog code of a particular module. Verilog code is organized in such a way that child modules are instantiated as components within a higher-level parent module, forming a hierarchical tree structure of instances as illustrated in Fig. \ref{fig-cpu-tree}. We consider the hierarchical structure is an important part of the instance information $c$. More specifically, all the ancestor instances of the instances form a path and every direct child instances with their corresponding power consumption values are listed, which is represented in an indent-based tree form as show in the left part of Fig. \ref{fig-represent-model}. Note that we only provide the module names because including the code of all these modules as input would result in an excessively long length.

\subsubsection{Segmented Verilog Code} \label{sec-sample-verilog}

Using the full Verilog code for certain instances might pose a challenge due to excessive code length. To address this, we truncate overly long code into multiple segments, as illustrated by the green shaded areas on the right side of Fig. \ref{fig-represent-model}. 

Segments, while only representing a portion of the original code, are randomly sampled and concatenated to construct the input $x=(c,w)$. The key distinction among these inputs is the specific code segment, while all other contextual information and the corresponding power consumption remain constant. In machine learning terms, this approach is analogous to data masking, where the model is exposed to only partial information yet is trained to robustly and accurately predict predict the output. Notably, the sampled Verilog code segments are positioned at the end of the input prompt. When the overall input length necessitates truncation, the Verilog code information is the first to be reduced, minimizing its impact on the model's predictive accuracy.

Modules lacking child modules, often referred to as leaf modules, typically possess concise Verilog code and are commonly instantiated multiple times across a design. We argue that these modules are typically too fine-grained and, individually, contribute negligibly to overall power consumption, thus attracting minimal designer attention. Consequently, leaf modules with code shorter than a defined threshold are treated as an integral part of their respective parent modules. Their Verilog code is then concatenated with that of their respective parent modules before segmentation.

\subsubsection{Workload and Other Parts}\label{sec-workload}

Currently, our power prediction methodology utilizes simple C++ programs as workloads, with their code manually selected to serve as input for the LLM. The extension to handling larger or more diverse programs is reserved for future research.

We employ a unified instruction format to prompt the LLM for power prediction. The prompt explicitly directs the LLM to sum the power consumption of all child modules' power consumption and to ensure the predicted value is not less than the sum. To provide the LLM with an understanding of the overall CPU context, microarchitecture parameters are also included as part of the input $x=(c,w)$ in JSON format, depicted on the middle-left side of Fig. \ref{fig-represent-model}.

Finally, inspired by the work of Gruver et al. \cite{Gruver2023zsfLargeLM}, we introduce an encoding method for floating-point power numbers to enhance the LLM's numerical stability. Specifically, decimal points are omitted by fixing the precision, transforming numbers like $1.234$e$-3$ to $1234$e$-6$. 

\subsection{Fine-tuning} \label{sec-fine}

We choose a pre-trained open source model,
LLaMa 3.1 8B-Instruct \cite{Dubey2024Thellama3} as the foundation model of BigPower, which is a famous light weight model and is well supported by the society. Our fine-tuning process is implemented using the LLaMa-Factory \cite{zheng2024llamafactory}. 

We use the supervised fine-tuning and detailed hyper-parameters setup will be discussed in experiments setup Section \ref{sec-exp-setup}.

To uniquely identify power prediction results for different instances, we incorporated an \texttt{id} property as meta-information within our dataset. To accommodate this, a minor adaptation is made to LLaMA-Factory's output handling, specifically to bypass the default processing of this \texttt{id} meta-property. This modification solely pertains to metadata handling and does not alter LLaMA-Factory's fundamental training or inference processes.

\subsection{Predicting} \label{sec-predicting}

The input $x$ for a given instance requires knowledge of the power consumption of its child modules. During the training phase, the ground truth power consumption for all modules (both parent and child) is known. This allows us to circumvent the inherent inter-module dependencies during training, as the model has direct access to the required child module power values.

For the prediction process, we propose a hierarchical "bottom-up" approach. In this methodology, module instances are systematically classified according to their level within the CPU's hierarchical tree structure (illustrated in Fig. \ref{fig-cpu-tree}). Power consumption estimates are then progressively computed, starting from the lowest-level modules (the "bottom") and proceeding upward to their respective parents, until the entire CPU's power is estimated.

Numerical stability is a concern when employing LLMs for regression tasks, particularly when dealing with floating-point predictions. It is clear that the digits in the exponent portion of a floating-point number influence its magnitude and, consequently, the prediction's overall accuracy, far more than the fractional part. To enhance the robustness of our LLM's predictions, especially given the inherent challenges of directly predicting precise floating-point values, we implement several strategies.

\textbf{Ensemble prediction via segmentation}: As detailed in Section \ref{sec-sample-verilog}, a single long Verilog code from instance is segmented, generating multiple distinct inputs for the LLM. Each segment yields an power consumption prediction for the same instance. It should be noted that during training, the ground truth labels correspond to the power of the entire module rather than that of individual segments, as per-segment power data is unavailable. Consequently, every segment within the same module is trained to predict the total module power, essentially estimating the whole from various constrained perspectives.
To mitigate variance and improve stability, the median of these multiple predictions is then taken as an ensemble result. This is inspired by ensemble learning, which is widely applied in machine learning. In this approach, each learner makes predictions about the whole based on partial features, and then the predictions from all learners are integrated (through voting for classification or averaging for regression). 

\textbf{Outlier handling and magnitude correction}: We also integrate a mechanism for addressing anomalous predictions. During the prediction process, we calculate the average power consumption and its standard deviation across all modules in the training set. It has been observed that the LLM occasionally predicts power values with significant errors in the exponent (magnitude), particularly when the estimation deviates substantially from the historical average. To counter these numerical instabilities, such outlier predictions are identified and subsequently replaced with the calculated average power consumption. 

\section{Experiment Settings} \label{sec-exp-setup}

\subsection{EDA-flow} \label{sec-eda-flow} Our work utilizes the XiangShan Nanhu version of the CPU design. All Verilog code is generated using the official Chisel framework, and all comments and debug information are systematically removed to ensure a clean input for subsequent processing (e.g., LLM ingestion). The gate-level netlist is then synthesized using Synopsys Design Compiler, while simulations are performed with Synopsys VCS. Finally, detailed power reports are generated using Synopsys PrimeTime PX. 
We have labeled the power consumption for 400 CPU configurations. Due to hardware resource limitations, fine-tuning with the entire dataset would be excessively time-consuming. Therefore, in this paper, we did not use the full dataset when validating our method. We randomly select 20 configurations as training data and 4 as test data in our experiments. We plan to open-source the complete labeled dataset. 

\subsection{Model tuning} The experimental setup comprise four NVIDIA L40S GPUs, each with 48GB of ECC memory. During the fine-tuning of the LLM, a learning rate of $2\times10^{-6}$ is used with a cosine scheduler and no warmup steps. Fine-tuning is limited to a single epoch on the full training dataset. To mitigate memory and computational resource limitations, BFloat16 (BF16) quantization is employed. We used a global batch size of 4 (one sample per GPU) with 32 gradient accumulation steps. During fine-tuning, a partial training strategy is adopted where only the last 10 layers of the LLM are updated, while all remaining trainable modules are frozen, including the default tokenizer and embedding layer of original LLaMA 3.1. For efficient memory management and accelerated training, we leverage DeepSpeed's \cite{Rasley2020DeepSpeedSO} Zero Redundancy Optimizer (ZeRO) Stage 2 \cite{Rajbhandari2019ZeROMO} with CPU offloading. The maximum sequence length is set at 8192 tokens. Sequences exceeding this limit will be truncated via the framework's dynamic length allocation strategy.  Model hyper-parameters are chosen based on common settings found in similar LLM fine-tuning tasks and not subjected to comprehensive optimization, given our resource constraints.

\subsection{Metrics}

Power modeling constitutes a regression problem. In evaluating the accuracy of modeling outcomes, we employ two commonly adopted metrics: the Mean Absolute Percentage Error (MAPE) and the Coefficient of Determination ($R^2$). 

\begin{equation}
    \mathrm{MAPE} = \frac{1}{n}\sum_{i}^{n}|\frac{\hat{y_{i}}-y_{i}}{y_{i}}| \times 100\%    
\end{equation}

\begin{equation}
    R^2 = 1- \frac{\sum_i^n(\hat{y_i}-y_i)^2}{\sum_i^n(y_i-\overline{y})^2}    
\end{equation}
where $y_i$ is the ground truth, $\hat{y_i}$ is the prediction, and $\overline{y}$ is the average of ground truth derived from the commercial EDA-flow detailed in  in Section \ref{sec-eda-flow}.
MAPE is calculated based on the normalized difference (often expressed as a percentage) between the actual power values and their corresponding predicted values. A lower MAPE signifies a smaller modeling error, indicating more accurate predictions.
$R^2$ measures the proportion of the variance in the dependent variable that is explainable by the independent variables within a regression model. A high $R^2$ value denotes a strong correlation between the variables. The maximum value of $R^2$ is 1, which indicates that the predictive model perfectly matches the true values.

\section{Main Results}\label{sec-main-result}
In this section, we test the generalization ability of BigPower and compare it with traditional models.

\begin{figure*}[!htbp]
\centering
\includegraphics[width=2\columnwidth]
{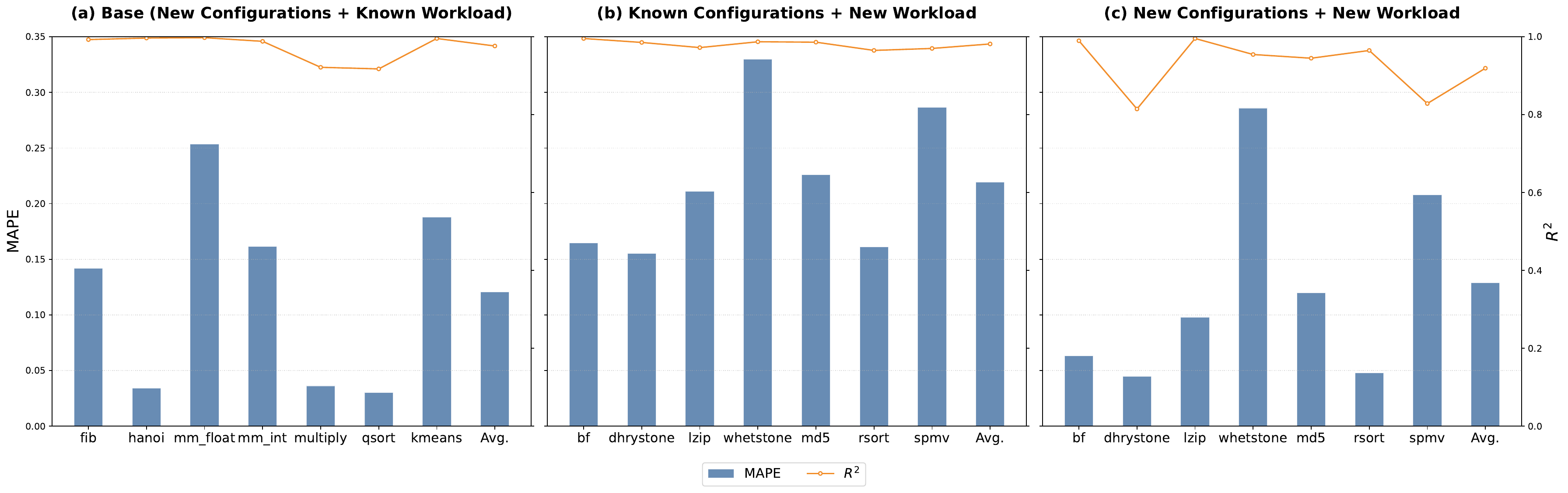}
\caption{Validation of BigPower. Comparison of MAPE (bars) and $R^2$ (lines) across varying configurations and workloads. Sub-labels on the X-axis denote specific benchmarks, with "Avg." representing the mean performance across the test. }
\label{fig-generalization-results}
\end{figure*}

\subsection{Generalization Across Configurations and Workloads}

To evaluate the performance on unseen microarchitecture configurations and workloads, we test the generalization ability of BigPower under new CPU configurations, new workloads, and combinations of both. The results, comprising MAPE and $R^2$ scores, are summarized in Fig. \ref{fig-generalization-results}.

\subsubsection{New Configurations, Known Workloads}
First, we evaluate the model on four unseen configurations using the seven workloads included in the fine-tuning set. As shown in Fig. \ref{fig-generalization-results}(a), the model achieves an average MAPE of 12.09\% and a strong $R^2$ of 0.97. While performance varies by workload ranging from 3.02\% to 25.36\%, the consistently high $R^2$ values (0.91--0.99) indicate that BigPower effectively captures the power scaling trends across different hardware parameters.

\subsubsection{Known Configurations, New Workloads} 
Second, we test the model's capability on seven entirely new workloads across known configurations.

\begin{enumerate}
  \item bf:  Brainfuck interpreter.
  \item dhrystone: A synthetic benchmark designed to measure integer arithmetic and string manipulation performance.
  \item lzip: Data compression utility that employs the Lempel-Ziv-Markov chain algorithm (LZMA)
  \item whetstone: A synthetic benchmark focused on floating-point arithmetic operations.
  \item md5: An implementation of the MD5 cryptographic hash function
  \item rsort: Radix sort algorithm.
  \item spmv: Sparse Matrix-Vector Multiplication.
\end{enumerate} 

As shown in Fig. \ref{fig-generalization-results}(b), the average MAPE increases to 21.94\%, with an average $R^2$ of 0.94. This increase in error is likely attributable to the relatively sparse diversity of the training workload. 

\subsubsection{New Configurations and New Workloads} 
Finally, we consider the most challenging scenario: predicting power for new workloads on new configurations. As shown in Fig. \ref{fig-generalization-results}(c), BigPower achieves a surprisingly average MAPE of 12.90\% and an average $R^2$ of 0.92. Although the average error remains low, we observe a notable decline in $R^2$ for specific benchmarks, such as whetstone and spmv. The decline in $R^2$ indicates a diminished capacity to track underlying trends compared to the previous two scenarios. 

Nevertheless, in all three evaluation settings, BigPower achieves an overall average MAPE of 15.64\% and $R^2$ of 0.94, demonstrating its effectiveness in fine-grained module power prediction without the need for simulation. 

\subsection{Comparison with Traditional Models}

It is difficult to perform direct comparisons with other power models across different target designs and application scenarios. To achieve the fine-grained, module-level power prediction demonstrated in our work, frameworks such as Primal \cite{2019ZhouPRIMALPowerInference} and Apollo \cite{Xie2021APOLLOAA} would require constructing and training individual models for each of the nearly 5,000 module instances within the XiangShan processor. Furthermore, Primal and Apollo operate at the RTL level. While GRANNITE \cite{2020ZhangGRANNITEGraphNeural} offers transferable predictions, it necessitates both RTL simulation and logic synthesis to get the dynamic toggle rates of the registers and the graph representation of the gate-level netlist. For large-scale circuits like XiangShan, the resulting graph size becomes prohibitively massive.

In prior works \cite{lee2006accurate, lee2007illustrative}, design parameters are used to perform regression modeling for design space exploration. Because these methods also do not utilize event characteristics obtained through dynamic simulation, they are unable to model varying benchmark behaviors and lack the capability for fine-grained power prediction.

To enable fine-grained power consumption prediction by traditional regression models, avoiding the need for features generated from dynamic simulations, we adopt a one-hot encoding scheme for both workloads and modules. The resultant input vectors consist of these one-hot encodings alongside system configuration parameters. A key challenge in encoding modules within the XiangShan architecture is that a single module code can correspond to a variable number of instantiated physical instances depending on the configuration. For the purpose of our encoding, all modules with the identical name are grouped and treated as the same unit. To account for this variability, our approach involves one-hot encoding the module's name only, with the prediction targeting the average power consumption across all its corresponding instances.

We experiment with different types of commonly used regression models, including Ridge, Lasso, SVR, Random forest, Gradient boost, Neural network.

Table \ref{tab:comp-tradition} presents the MAPE values for these traditional regression models and our BigPower. As can be seen, the two linear regression models perform extremely poorly on this task. While SVR, Random forest, Gradient boost, and Neural network perform slightly better than linear regression, their average MAPE values still range from 182.22\% to 449.62\%. Only our BigPower achieves satisfactory prediction results.

\begin{table*}[!htbp]
  \centering
  \caption{Comparison of MAPE (\%) with different models.}

    \begin{tabular}{
    >{\centering\arraybackslash}p{1.6cm}>{\centering\arraybackslash}p{1.8cm}>{\centering\arraybackslash}p{1.8cm}>{\centering\arraybackslash}p{1.8cm}>{\centering\arraybackslash}p{1.8cm}>{\centering\arraybackslash}p{1.8cm}>{\centering\arraybackslash}p{2cm}>{\centering\arraybackslash}p{1.6cm}}
    \hline
             & Ridge & Lasso & SVR   & Random Forest & Gradient Boost & Neural Network & Ours \\
          \hline
    fib &  11334.31 & 73695.22 & 206.22 & 278.49 & 558.07 & 418.66 & 14.21 \\
              \hline
    hanoi &    6111.65 & 52328.31 & 130.17 & 194.79 & 308.77 & 339.22 & 3.43 \\
              \hline
    mm\_float    & 9398.31 & 73904.33 & 185.94 & 280.59 & 465.54 & 425.52 & 25.36 \\
              \hline
    mm\_int    & 11588.88 & 74227.57 & 238.73 & 284.36 & 548.95 & 403.95 & 16.18 \\
              \hline
    multiply    & 8009.40 & 69553.33 & 170.59 & 262.65 & 498.59 & 420.13 & 3.64 \\
              \hline
    qsort    & 9211.21 & 67538.09 & 166.66 & 256.29 & 533.02 & 478.78 & 3.02 \\
              \hline
    kmeans    & 5362.24 & 45925.41 & 177.22 & 170.55 & 234.42 & 343.69 & 18.79 \\
             \hline
    \textbf{Average}    & 8716.57 & 65310.32 & 182.22 & 246.82 & 449.62 & 404.28 & 12.09 \\
            \hline
    \end{tabular}%
  \label{tab:comp-tradition}%
\end{table*}%

\section{Ablation Study}

This section presents several ablation experiments to verify the effectiveness of the strategies employed during data construction and the prediction process.

\subsection{Ablation Study for Dataset Construction}

We conduct preliminary ablation studies to determine the optimal strategy for constructing the training dataset. For rapid iteration, several experiments utilize a simplified setup. The test set contains randomly sampled  $7\times10^{5}$ pairs, with $1\times10^{5}$ pairs drawn from each of the seven distinct workloads. The test dataset use CPU configurations that are unseen during training. For the ablation study, test and training datasets are generated and evaluated pair-wise, in order to explore the effects of different dataset construction methodologies. Notably, the ground-truth power values of child modules are provided directly in the input prompt for this ablation study. 

\subsubsection{Impact of Dataset Size} We first evaluate the correlation between training data volume and model performance. As illustrated in Fig. \ref{fig-cmp-dataset}(a), prediction accuracy improves consistently as the dataset scales. The model trained on the "full" dataset significantly outperforms smaller subsets, achieving the lowest MAPE. The evaluated configurations are defined as follows:

\begin{enumerate}
    \item small: A random sample of $7\times 10^{5}$ input pairs.
    \item medium: A larger set of  $7\times10^{6}$ pairs, sampled evenly from 7 distinct workloads.
    \item large: An even larger set of $1.4\times10^{7}$ pairs, sampled evenly from 7 distinct workloads.
    \item full: One complete training epoch on the entire dataset.
\end{enumerate}

\subsubsection{Impact of Prompt Composition}

\begin{figure}[!htbp]
\centering
\includegraphics[width=1.0\columnwidth]
{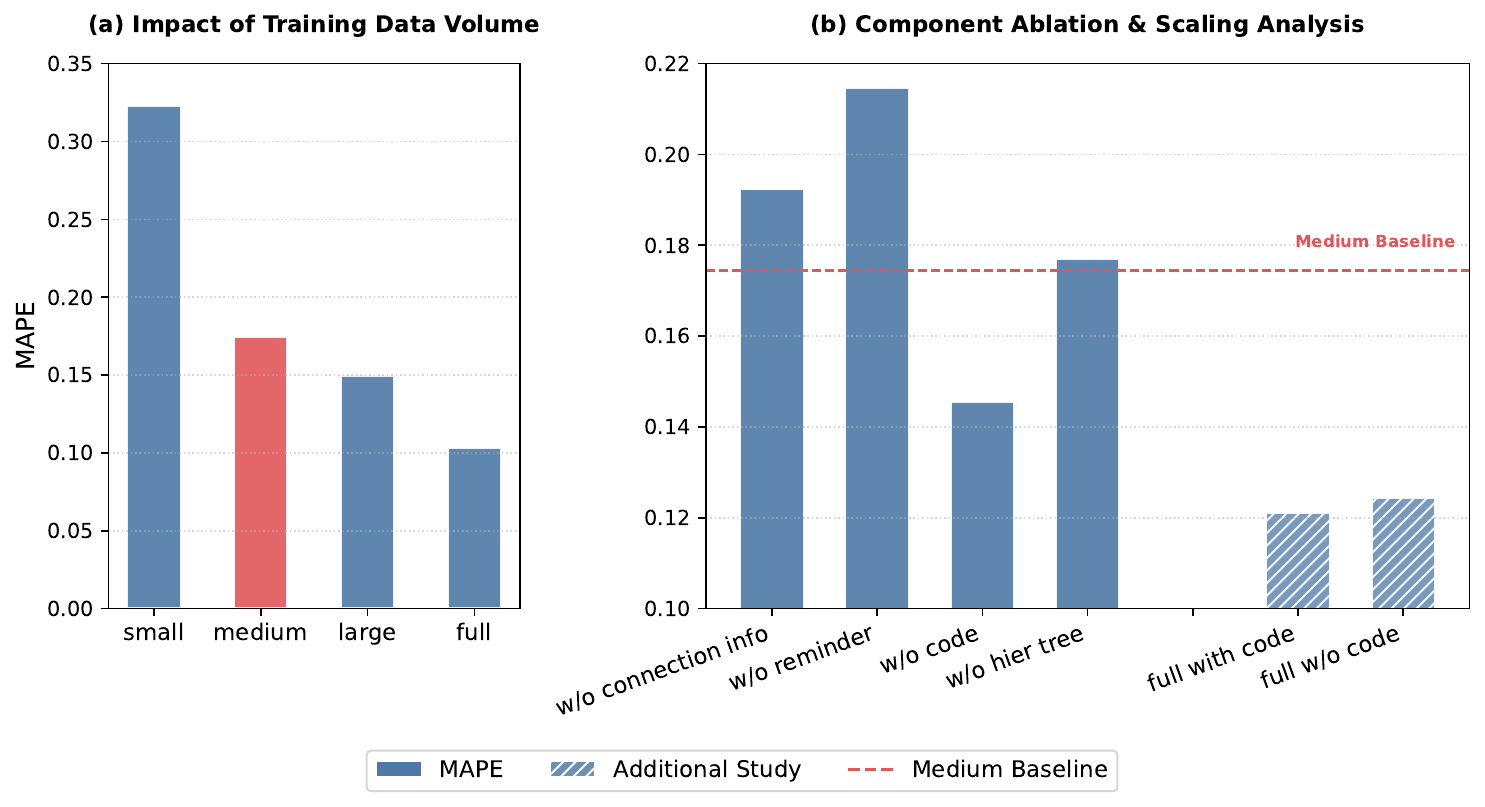}
\caption{Evaluation of dataset construction strategies. (a) Impact of training data scale on model accuracy, showing a consistent reduction in MAPE as the dataset scales. (b) Ablation of prompt components on the medium dataset (left), followed by a comparative validation of the Verilog code impact at full scale (right). The red dashed line represents the baseline performance of the standard medium dataset configuration.}
\label{fig-cmp-dataset}
\end{figure}

We further investigate the contribution of individual components within the LLM prompt. These experiments utilize the medium dataset to balance training efficiency with performance insights. The results are summarized in Fig. \ref{fig-cmp-dataset}(b). We examine the following variations:

\begin{enumerate}
    \item w/o connection info: Omits explicit information about sibling module connectivity.
    \item w/o reminder: Removes instructional cues and reminders from the prompt.
    \item w/o code: Excludes Verilog code while retaining all other prompt elements.
    \item w/o hier tree: Removes the hierarchical tree structure
\end{enumerate}

Using the medium configuration as the baseline (indicated by the red dashed line), most ablation variants exhibit inferior performance (higher MAPE). A notable exception is the w/o code variant, which appears to outperform the baseline in the simplified setup. We suppose that Verilog code, as a form of unstructured text, is intrinsically information-sparse. Consequently, the model can only effectively leverage such information when supported by a sufficiently large training dataset. To investigate this further, we conducted a full-scale training run without code, evaluated using the final bottom-up paradigm detailed in Section \ref{sec-predicting}. As shown in the right-hand portion of Fig. \ref{fig-cmp-dataset}(b), our proposed method (with code) ultimately achieves a lower MAPE when scaled to the full dataset, justifying its inclusion in the final architecture.

\begin{figure}[!htbp]
\centering
\includegraphics[width=1.0\columnwidth]
{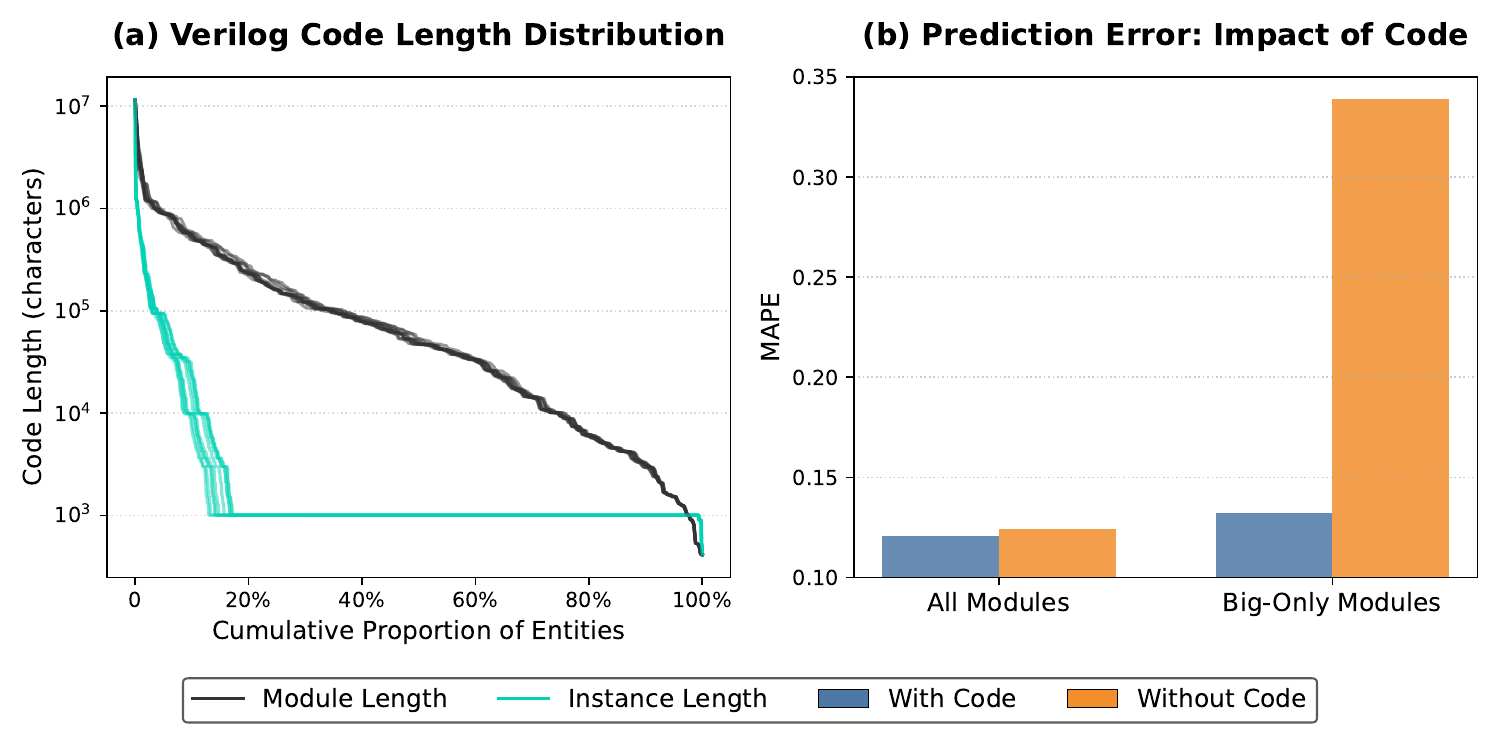}
\caption{Analysis of Verilog code complexity. (a) Cumulative distribution of Verilog code length (after cleaning) for modules and their corresponding instances. Each line represents a unique CPU configuration. (b) Comparison of prediction error between the full proposed model and the "without code" variant.}
\label{fig-mlen}
\end{figure}

To further investigate the impact of Verilog code on model performance, we characterized the code length distribution across different modules in XiangShan. As illustrated in Fig. \ref{fig-mlen}(a), the majority of  instances in XiangShan are relatively small in terms of the Verilog code length. Intuitively, while the power characteristics of simple modules may be easily captured via structural information, complex modules described by extensive code segments are likely to derive greater benefit from our code segment strategy.

To test this, we isolated a subset of the "largest" modules (the top 10\% by code length) to compare model performance. As shown in Fig. \ref{fig-mlen}(b), our proposed method achieves a robustly low MAPE across both the full test set and the big-module subset. In contrast, the performance of the "without code" variant degrades significantly when restricted to these complex modules, experiencing a sharp increase in MAPE. This confirms that the inclusion of Verilog code is critical for modeling the power consumption.

\subsection{Ablation Study for Predicting}

We conduct an ablation study to validate  the effectiveness of the bottom-up prediction strategy detailed in Section \ref{sec-predicting}. As discussed in Section \ref{sec-sample-verilog}, instances with Verilog code exceeding the model's context length are segmented, necessitating an ensemble strategy to consolidate their multiple prediction outputs. Given that our recursive bottom-up approach has the potential to allow errors to propagate through the hierarchy, the choice of ensemble method is critical. The following configurations are evaluated:

\begin{enumerate}
    \item Base (Proposed Method): The standard recursive approach. For segmented instances, the median of all segment predictions is used as the module's power value. This value then serves as the child-module input for the parent's power prediction.
    \item w/o bottom-up: The recursive approach is disabled; power values for all child modules are fixed at default value of zero in the input as a placeholder for missing hierarchical information. 
    \item First Element: A baseline that uses only the prediction from the first code segment.
   \item Random: A baseline that randomly selects the prediction from one of the segments.
\end{enumerate}

\begin{figure}[!htbp]
\centering
\includegraphics[width=1\columnwidth]
{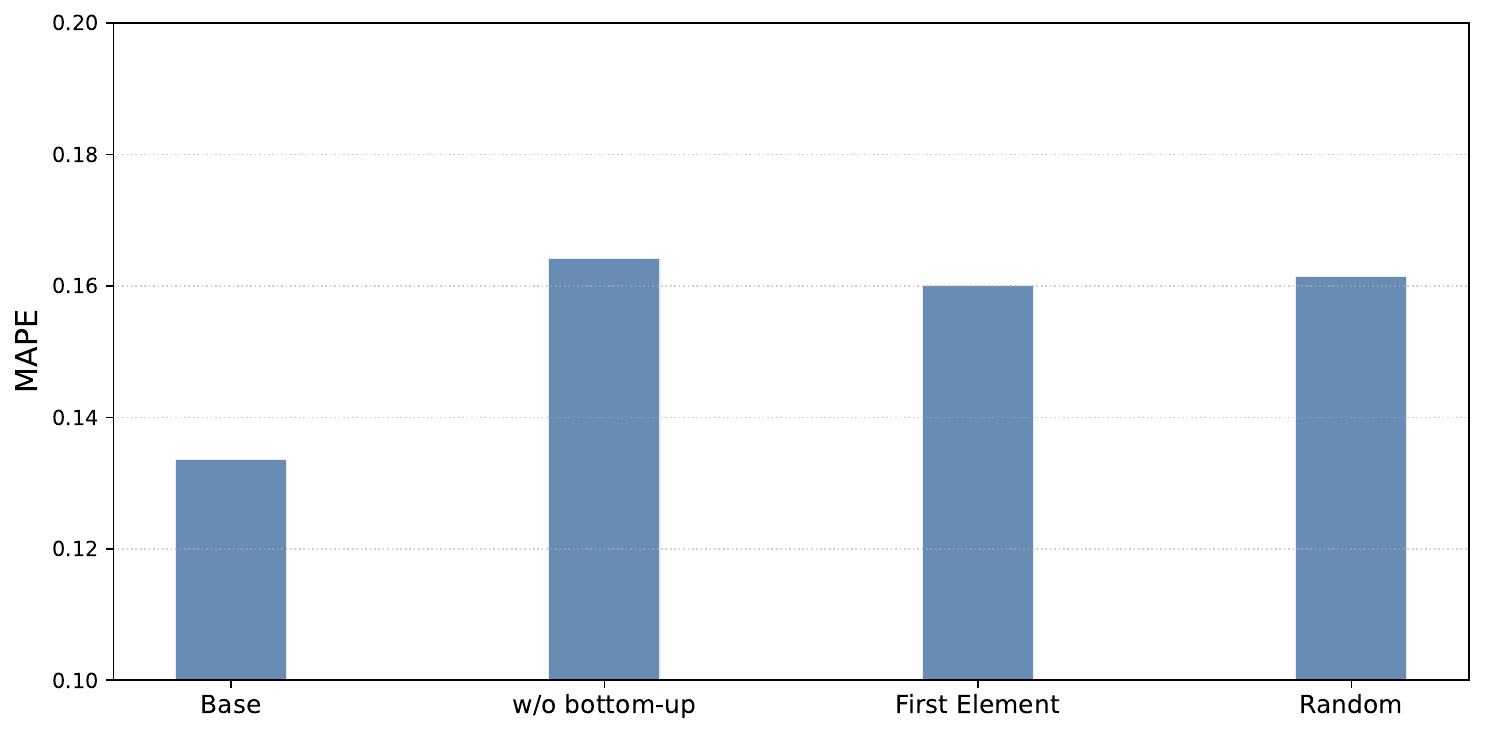}
\caption{Impact of hierarchical prediction and segment ensemble.}
\label{fig-cmp-predict}
\end{figure}

As shown in Fig. \ref{fig-cmp-predict}, the Base method achieves the lowest MAPE. The w/o bottom-up variant is inferior to all recursive approaches. We attribute this partially to the model encountering out-of-distribution (OOD) data; as the LLM is not trained on inputs where all child modules have zero power, leading to poor generalization in this scenario.  Furthermore, the First Element and Random methods are inherently arbitrary and exhibit less numerical stability than our median-based approach, resulting in higher overall error.

\section{Case Study}

In this section, we show the practical advantages of BigPower through two use cases. 1) Hierarchical power breakdown: our ability to perform fine-grained, module-level predictions allows for a layer-by-layer diagnostic of specific designs. 2) Design change sensitivity analysis: we can clearly contrast power fluctuations resulting from architectural modifications. Crucially, these applications eliminate the requirement for simulation. By leveraging sufficient GPU resources, BigPower could enable massive parallel prediction across modules, thereby accelerating the design iteration cycle.

\subsection{Hierarchical Power Breakdown}
\label{sec:case-study 1}

\begin{figure*}[!htbp]
\centering
\includegraphics[width=2\columnwidth]
{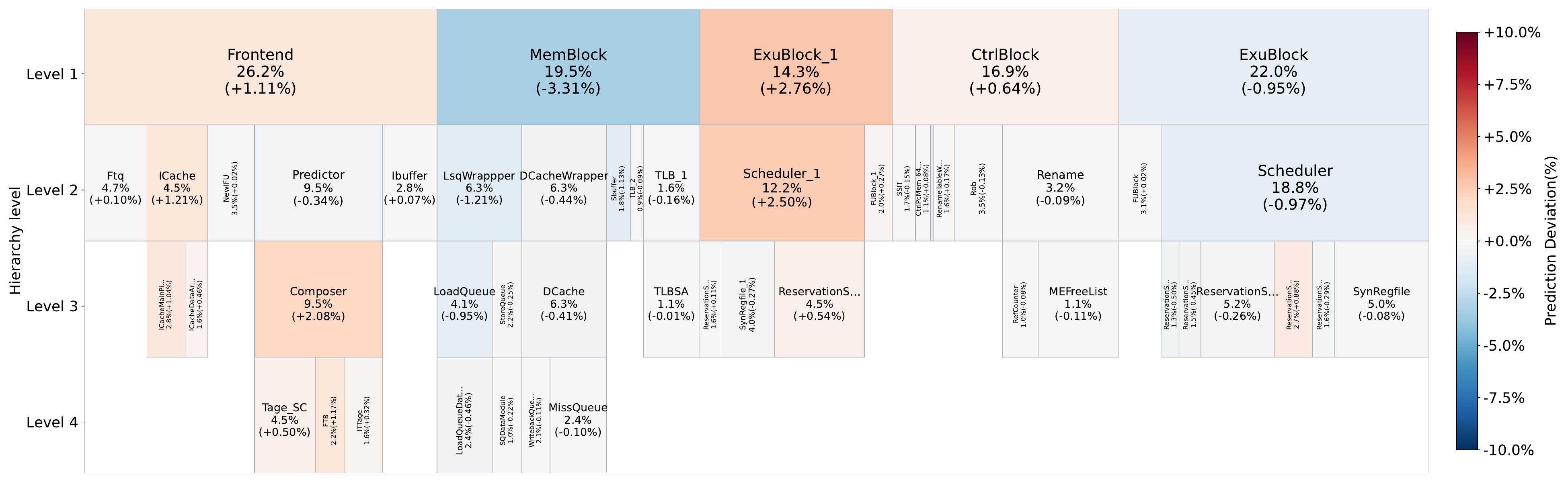}
\caption{Hierarchical power breakdown. For visual clarity, the root and negligible sub-modules are omitted. Each block displays the module name, its predicted power share in current hierarchical level, and relative error $\Delta_P$ (in brackets). Color intensity represents prediction error, with red (+) for overestimation and blue (-) for underestimation.}
\label{fig-breakdown}
\end{figure*}

In contrast to most existing works, BigPower predicts power consumption directly at the source code level, facilitating a comprehensive top-down analysis. This granularity enables researchers to pinpoint and optimize power hot spots. Fig. \ref{fig-breakdown} illustrates the prediction error distribution across the architectural hierarchy with a specific configuration (Configuration 1, $C_1$) on the workload Hanoi.

We define the relative error as:

\begin{equation}
    \Delta_P=\frac{P_{module\_predict}}{P_{root\_predict}}-\frac{P_{module\_label}}{P_{root\_label}}
\end{equation}

This metric characterizes the model's accuracy in identifying the relative power proportion of each component. In the visualization, the color intensity represents the error magnitude, where red hues indicate an overestimation of a module's power share and blue hues indicate an underestimation.

Using the \texttt{XSCore} as the root module, the visualization reveals a consistent "near-white" color profile across all hierarchical levels. This signifies that BigPower maintains high fidelity in preserving spatial power locality, ensuring that the predicted power profile closely mirrors the ground-truth distribution across functional units. Minor deviations occur in \texttt{Scheduler\_1} and \texttt{ExuBlock\_1}, where the model slightly overestimates power contributions by +2.50\% and +2.76\%, respectively. Notably, \texttt{ExuBlock\_1} serves as the floating-point unit in the XiangShan architecture. Although the selected workload (Hanoi) is primarily a fixed-point program, the early-stage XiangShan design features relatively immature clock gating. Consequently, BigPower correctly captures a significant power share in modules that would otherwise be expected to remain idle or "dimmed" in a more power-optimized implementation.

\subsection{Design Change Sensitivity Analysis}
\begin{figure*}[!htbp]
    \centering
    \includegraphics[width=2\columnwidth]
    {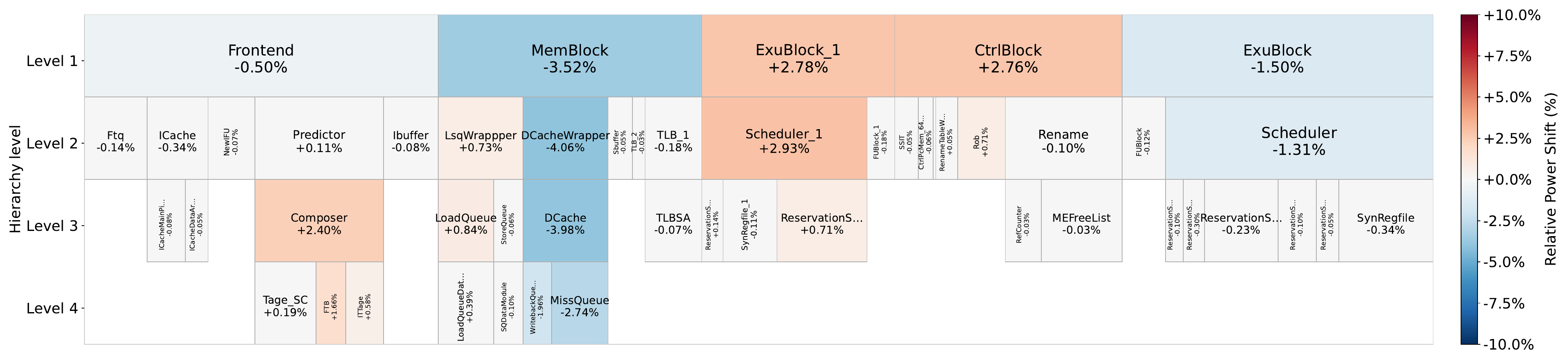}
    \caption{
    Design change sensitivity analysis. Each block displays a module's name and its relative power shift ($\Delta_{shift}$) between $C_1$ and $C_2$. Color intensity reflects the magnitude of change: red (+) represents an increased power footprint in $C_1$, while blue (-) represents a decrease.
    }
    \label{fig-cmpcfg}
\end{figure*}

To demonstrate the diagnostic utility of BigPower, we perform a comparative analysis between two distinct processor configurations. Fig. \ref{fig-cmpcfg} provides a fine-grained view of how architectural parameter tuning impacts the core's power profile.

In this case, we define the predicted power shift ($\Delta_{shift}$) as the difference in the predicted power proportion of a module between Configuration 1 ($C_1$) and Configuration 2 ($C_2$):

\begin{equation}
    \Delta_{shift} = \left( \frac{P_{module\_predict}}{P_{root\_predict}} \right)_{C_1} - \left( \frac{P_{module\_predict}}{P_{root\_predict}} \right)_{C_2}
\end{equation}

The parameter variations between $C_1$ and $C_2$ are summarized in Table~\ref{tab:cfg-diff}. Notably, $C_1$ features fewer Miss Status Holding Register (MSHR) entries than $C_2$. This architectural reduction is apparent in Fig. \ref{fig-cmpcfg}, where the \texttt{DCache} and \texttt{DCacheWrapper} blocks are highlighted with deep blue shading. Conversely, $C_1$ employs a larger \texttt{LoadQueue} (64 entries) compared to $C_2$ (48 entries). BigPower captures this overhead, represented by the light red tint in the \texttt{LoadQueue} block. Our results indicate that TLB entry count has a negligible impact on power distribution for this specific workload. It should be noted that the L3 Cache is excluded from Fig. \ref{fig-cmpcfg}, as the visualization is restricted to the power distribution within the \texttt{XSCore}. Furthermore, configuration changes may indirectly affect certain modules, e.g.  \texttt{ExuBlock\_1} and \texttt{CtrlBlock}.

\begin{table}[!htbp]
  \centering
  \caption{Parameter variations in configurations $C_1$ and $C_2$.}

    \begin{tabular}{ccccc}
    \hline
          & ldq\_entry & dcache\_tlb & dcache\_mshr & l3\_size (MB)\\
          \hline
    $C_1$ & 64    & 32    & 32    & 16 \\
    $C_2$ & 48    & 64    & 64    & 32 \\
    \hline
    \end{tabular}%

  \label{tab:cfg-diff}%
\end{table}%

\section{Limitations}

Although we have effectively predicted power consumption using LLMs, there remain certain limitations. First, cycle-accurate power prediction remains challenging. Predicting per-cycle power consumption without dynamic simulation is an extremely challenging task. It requires predicting power fluctuations from the start to the end of program execution.
Second, we do not account for back-end physical implementation. Consistent with most power models, our current approach focuses on evaluating the power performance of CPU front-end of power flow.
Third, the complexity of workloads is currently limited. Labeling power consumption using PTPX is time-consuming, making it nearly impossible for complex workloads. This prevents us from obtaining true power consumption values for these complex workloads through simulation, a challenge faced by power modeling research.

\section{Conclusion}

This paper presents BigPower, an LLM-based hierarchical framework for fine-grained module-level power estimation from source-level CPU design information. By leveraging workload characteristics, architectural hierarchy, module connectivity, configuration parameters, and RTL code representations, BigPower estimates module-level power consumption without requiring additional cycle-level simulation during inference. BigPower is trained on an extensive module-level labeled dataset spanning diverse workloads and CPU configurations of the XiangShan processor family. To address hierarchical dependency and long-context challenges, BigPower employs a structured representation combining global and local CPU context together with a bottom-up hierarchical inference strategy and ensemble aggregation. Experimental results demonstrate practical within-family generalization across unseen CPU configurations and workloads in XiangShan. Overall, BigPower provides an efficient and practical framework for fine-grained module-level power estimation during CPU development.

While BigPower currently focuses on within-family module-level estimation and does not model physical back-end effects, our findings suggest that source-level representations combined with hierarchical architectural context provide a promising direction for practical CPU power estimation. Future work includes extending workload diversity, improving cross-architecture transferability, and exploring finer temporal granularity.


\clearpage
\bibliographystyle{ACM-Reference-Format}
\bibliography{sample-base}

\end{document}